\documentclass[a4paper,fleqn,usenatbib,useAMS]{mnras}

\usepackage{graphicx}	
\usepackage{amsmath}	
\usepackage{amssymb}	
\usepackage{multicol}        
\usepackage{bm}		
\usepackage{pdflscape}	
\usepackage{flushend}

\usepackage[T1]{fontenc}
\usepackage{ae,aecompl}

\usepackage{newtxtext,newtxmath}

\title[Jet Kinematics of the Quasar 4C $+$21.35 from KaVA Observations]{Jet Kinematics of the Quasar 4C $+$21.35 from Observations with the KaVA Very Long Baseline Interferometry Array}

\author[Lee et al.]
{
Taeseok Lee$^{1}$\thanks{Contact e-mail: \href{mailto:tlee@astro.snu.ac.kr}{tlee@astro.snu.ac.kr}},
Sascha Trippe$^{1}$\thanks{Corresponding author,  \href{mailto:trippe@astro.snu.ac.kr}{trippe@astro.snu.ac.kr}},
Motoki Kino$^{2,3}$,
Bong Won Sohn$^{4}$,
Jongho Park$^{1}$,
\newauthor
Junghwan Oh$^{4}$, Kazuhiro Hada$^{5,10}$, Kotaro Niinuma$^{6}$, Hyunwook Ro$^{4,7}$, Taehyun Jung$^{4}$,
\newauthor
Guan-Yao Zhao$^{4}$, Sang-Sung Lee$^{4}$, Juan-Carlos Algaba$^{1,16}$, Kazunori Akiyama$^{8}$,
\newauthor
Kiyoaki Wajima$^{4}$, Satoko Sawada-Satoh$^{9}$, Fumie Tazaki$^{5}$, Ilje Cho$^{4,11}$, Jeffrey Hodgson$^{4}$,
\newauthor
Jeong Ae Lee$^{1}$, Yoshiaki Hagiwara$^{12}$, Mareki Honma$^{3,10}$, Shoko Koyama$^{13}$, Tao An$^{14,15}$,
\newauthor
Yuzhu Cui$^{3}$, Hyemin Yoo$^{7}$, Noriyuki Kawaguchi$^{14}$, Duk-Gyoo Roh$^{4}$, Se-Jin Oh$^{4}$,
\newauthor
 Jae-Hwan Yeom$^{4}$, Dong-Kyu Jung$^{4}$, Chungsik Oh$^{4}$, Hyo-Ryoung Kim$^{4}$, Ju-Yeon Hwang$^{4}$, 
\newauthor
Do-Young Byun$^{4}$, Se-Hyung Cho$^{4}$, Hyun-Goo Kim$^{4}$, Hideyuki Kobayashi$^{5}$,
\newauthor
Katsunori M. Shibata$^{3,5}$, Zhiqiang Shen$^{14,15}$, Wu Jiang$^{14,15}$, Jee Won Lee$^{4}$
\\ 
\\ 
$^{1}$Department of Physics and Astronomy, Seoul National University, Gwanak-gu, Seoul 08826, Korea \\
$^{2}$Kogakuin University of Technology \& Engineering, Academic Support Center, 2665-1 Nakano, Hachioji, Tokyo 192-0015, Japan \\
$^{3}$National Astronomical Observatory of Japan, 2-21-1 Osawa, Mitaka, Tokyo 181-8588, Japan \\
$^{4}$Korea Astronomy and Space Science Institute, Yuseong-gu, Daejeon 34055, Korea \\
$^{5}$Mizusawa VLBI Observatory, National Astronomical Observatory of Japan, Osawa, Mitaka, Tokyo 181-8588, Japan \\
$^{6}$Graduate School of Sciences and Technology for Innovation, Yamaguchi University, Yoshida 1677-1, Yamaguchi, Yamaguchi 753-8512, Japan \\
$^{7}$Department of Astronomy, Yonsei University, 134 Shinchon-dong, Seodaemun-gu, Seoul, 120-749, Republic of Korea \\
$^{8}$Massachusetts Institute of Technology, Haystack Observatory, 99 Millstone Road, Westford, MA 01886, USA \\
$^{9}$Graduate School of Science and Engineering, Kagoshima University, 1-21-35 Korimoto, Kagoshima-shi, Kagoshima 890-0065, Japan \\
$^{10}$Department of Astronomical Science, The Graduate University for Advanced Studies (SOKENDAI), 2-21-1 Osawa, Mitaka, Tokyo 181-8588, Japan \\
$^{11}$Department of Astronomy \& Space Science, University of Science \& Technology, 217 Gajeong-ro, Daejeon, Republic of Korea \\
$^{12}$Toyo University, 5-28-20 Hakusan, Bunkyo-ku, Tokyo 112-8606, Japan\\
$^{13}$Institute of Astronomy \& Astrophysics, Academia Sinica, P.O. Box 23-141, Taipei 10617, Taiwan\\
$^{14}$Shanghai Astronomical Observatory, Chinese Academy of Sciences, 80 Nandan Road, Shanghai 200030, China\\
$^{15}$Key Laboratory of Radio Astronomy, Chinese Academy of Sciences, Nanjing 210008, China\\
$^{16}$Department of Physics, Faculty of Science, University of Malaya, 50603 Kuala Lumpur, Malaysia\\
}

\date{Last updated 2018 April 20; in original form 2018}
\pubyear{2018}
\begin{document}
\label{firstpage}
\pagerange{\pageref{firstpage}--\pageref{lastpage}}
\maketitle

{
\vspace{3ex}
\bf 
Abstract follows on next page
}

\clearpage 

\begin{abstract}
We present the jet kinematics of the flat spectrum radio quasar (FSRQ) 4C~$+$21.35 using time-resolved KaVA very long baseline interferometry array radio maps obtained from September 2014 to July 2016. During two out of three observing campaigns, observations were performed bi-weekly at 22 and 43~GHz quasi-simultaneously. At 22~GHz, we identified three jet components near the core with apparent speeds up to $(14.4\pm2.1)c$. The timing of the ejection of a new component detected in 2016 is consistent with a $\gamma$-ray flare in November 2014. At 43~GHz, we found four inner jet ($<$3 mas) components with speeds from $(3.5\pm1.4)c$ to $(6.8\pm1.5)c$. Jet component speeds tend to be higher with increasing distances from the core. We compared our data with archival Very Long Baseline Array (VLBA) data from the Boston University (BU) 43~GHz and the Monitoring Of Jets in Active galactic nuclei with VLBA Experiments (MOJAVE) 15.4~GHz monitoring programs. Whereas MOJAVE data and our data are in good agreement, jet speeds obtained from the BU Program data in the same time period are about twice as high as the ones we obtain from the KaVA data. The discrepancy at 43~GHz indicates that radio arrays with different angular resolution identify and trace different jet features even when the data are obtained at the same frequency and at the same time. The flux densities of jet components decay exponentially, in agreement with a synchrotron cooling time scale of $\sim$1 year. Using known electron Lorentz factor values ($\sim$9\,000), we estimate the magnetic field strength to be $\sim$1--3~$\mu$T. When adopting a jet viewing angle of $5^{\circ}$, the intrinsic jet speed is of order $0.99c$. 
\end{abstract}

\newcommand{\RNum}[1]{\uppercase\expandafter{\romannumeral #1\relax}}
\section{Introduction\label{sec:intro}}

Active Galactic Nuclei (AGN) are the most powerful persistent astronomical sources and are marked, among others, by the ejection of relativistic jets of matter that reach kiloparsec to megaparsec scales in size. The detailed mechanisms of ejection and collimation of these powerful jets are only partially understood, even though there has been progress in understanding them both observationally and theoretically in recent years \citep[see, e.g.,][]{Komissarov2009, boettcher2012, Mertens2016}. Generally, the interaction of accreted matter, magnetic fields, and either black hole rotation or accretion disc rotation is supposed to launch a jet from the immediate vicinity (a few Schwarzschild radii) of a galactic supermassive black hole \citep{Blandford1977, Blandford1982}.

According to the unified model of AGN \citep{Urry1995}, blazars are AGN whose jet axis is aligned with the line of sight within a few degrees. They are characterised by strong stochastic (red-noise type) variability across the whole electromagnetic spectrum, from $\gamma$-rays to the radio domain, on time scales from hours to years \citep[e.g.,][]{Trippe2011, Gupta2012, Park2014, Aleksic2011}. 

Commonly, blazars are divided into two groups: BL Lacertae objects (BLO) and flat spectrum radio quasars (FSRQ) according to spectral features \citep[e.g.,][]{Beckmann2012}. Recent works by \citet{Ghisellini2011} proposed a more physical distinction based on the difference in accretion rates: FSRQs and FR \RNum{2} have disc luminosities higher than $5 \times10^{- 4}$ of the Eddington luminosity, while BLOs and FR \RNum{1}s have disc luminosities lower than that.
Further works expanded the classification by including the radio-loud galaxies. \citet{Meyer2011} suggested that radio-loud AGN are divided by source power: low synchrotron peaked sources whose jets do not show significant velocity gradients and high synchrotron peaked sources whose jets do show such gradients. As both types of structure are observed in AGN with low kinetic jet power, they suggested that the jet structure in those sources is controlled by the accretion rate in Eddington units, with the dividing line between the two source types being located between $3\times10^{-3}$ and $10^{-2}$. \citet{Sbarrato2014} also used a sample of blazars and radio-galaxies and found a smooth transition from radiatively inefficient to efficient accretion discs. \citet{Giommi2012} concluded from their Monte-Carlo simulations that there are only two intrinsic blazar types, low-ionization sources consisting mostly of beamed Fanaroff-Riley type \RNum{1} (FR \RNum{1}) galaxies and high-ionization sources of mostly beamed FR \RNum{2} ones. \citet{Xiong2015} also found an anti-correlation between jet power and beam-corrected synchrotron peak frequency of blazars and radio galaxies. \citet{Hervet2016} approached the blazar classification using very-long-baseline-interferometry (VLBI) maps, resulting in three classes with different radio knot kinematics.

The nearby FSRQ 4C~$+$21.35 (PKS 1222+216), at a redshift of 0.433, is a compact lobe-dominated source \citep{Wang2004} and located at the J2000 position $\alpha=12$h24m54.4584s, $\delta=+21\deg22'46.389''$. This source displays super-luminal apparent jet motions with speeds ranging from $3c$ to $25c$ at milli-arcsecond scales as revealed by VLBI maps \citep{Jorstad2001, Jorstad2014, MOJAVE13, Jorstad2017}. A large jet knot was ejected in the early 2000s and has kept its way along a bent trajectory since then \citep{MOJAVE7}. The bent trajectory is very different from the path of a jet knot ejected earlier; indeed, the two knots have been moving into different directions \citep{MOJAVE10}.

4C~$+$21.35 also belongs to a small group of very-high-energy (VHE; $E>100$~GeV) detected FSRQs along with 3C 279, PKS 1510--089, PKS 1441+25, S3 0218+35, and PKS 0736+017 \citep{Cerruti2017}. 4C~$+$21.35 experienced two major $\gamma$-ray flares in April and June 2010. Multiple studies \citep{Aleksic2011, Tavecchio2011, Nalewajko2014, Ackermann2014} aimed at finding the origin of the $\gamma$-ray radiation and at understanding the emission processes. \citet{Jorstad2014} used multi-frequency data including Very Long Baseline Array (VLBA) 43~GHz and \emph{Fermi} Large Area Telescope (LAT) light curves and found a good correlation between $\gamma$-ray emission and radio kinematics. They showed that the $\gamma$-ray flare occurred when a newly ejected jet component moves through the core as well as a stationary feature near the core, which are assumed to be recollimation shocks. The time it took a moving knot to cross the stationary knot, and the epoch of the crossing, coincide with that of $\gamma$-ray flares. 

\begin{table}
\centering
\caption{Overview of our KaVA observations. \label{tab:obs}}
\begin{tabular}{llllcccc}

\hline
Year & Month & Day & Frequency [GHz] \\
\hline

2014 & Sep & 1 & 22 \\ 
	 &     &     14 & 22 \\ 
	 & Nov    &      4& 22 \\\\   
	 	 
2015	 & May    &3     & 22 \\ 
	 &     &4   &     43 \\ 
	 &     &  16 &     22 \\ 
	 &     &  17 &     43 \\\\  
	 
2016	&	Feb&	25&	22 \\ 
	&	&	26&	43 \\ 
	&	Mar&9&	22 \\ 
	&	&  10	&	43 \\ 
	&	&	20&	43 \\ 
	&	&	21&	22 \\
	&	Apr&	8&	22 \\
	&	&	9&	43 \\
	&	&	21&	22 \\
	&	&	22&	43 \\
	&	May&	3&	22 \\
	&	&	5&	43 \\
	&	&	23&	22 \\
	&	&	24&	43 \\
	&	Jun&	2&	43 \\
	& &	13	&22 \\
	&	&	15&43 \\
\hline
\end{tabular}
\end{table}

So far, the VHE emission from FSRQs is usually understood as originating from a compact emitting region residing just outside the broad-line-region (BLR). The location is inferred from the absence of VHE photon absorption by BLR clouds \citep{boettcher2016}, and the small size is inferred from the rapid variability on time scales of a few minutes, thus implying physical extensions of less than a few light minutes \citep{Tavecchio2011}. \citet{Ackermann2014} constructed a theoretical spectral energy distribution (SED) of the source using a one-zone leptonic model in which a small emission region (with a radius of $\sim10^{15}$~cm) resides outside the BLR. The resulting SED model, which is fitted to multi-frequency data, has a characteristic two-hump (``camel back'') structure, with the first hump (from radio to optical) being caused by synchrotron radiations and the second (from optical to $\gamma$-rays) being due to inverse Compton radiation. Synchrotron self-Compton (SSC) radiation supposedly dominates the spectrum in the X-ray regime, external Compton (EC) emission dominates the $\gamma$-ray spectrum. 

In this work, we aim at a detailed characterization of the kinematics of 4C~$+$21.35 as well as illuminating the connection between kinematics and $\gamma$-ray activity. Since 4C~$+$21.35 shows ongoing vigorous $\gamma$-ray activity and blazars are famous for their fast variability, persistent and frequent monitoring of the source structure is essential. We therefore studied the kinematics of the jet of 4C~$+$21.3 with bi-weekly radio interferometric mapping observations, providing an unprecedented density of data.

We used the Korean VLBI Network (KVN) and VLBI Exploration of Radio Astrometry (VERA) array (KaVA; for detailed information, see \citealt{Niinuma2014,An2016,Zhang2017,Cho2017,Kino2018}) at 22 and 43~GHz from 2014 to 2016. We follow up on an earlier study by \citet{Oh2015} who observed eight radio-bright AGN with KaVA over one year. We compare our results with archival VLBA data from the Boston University Blazar Monitoring program at 43~GHz and Monitoring Of Jets in Active galactic nuclei with VLBA Experiments (MOJAVE) at 15~GHz that approximately cover the time span of our observations.

Throughout the paper, we adopt a cosmology with $\Omega_{m}=0.27$, $\Omega_{\Lambda}=0.73$, and $\rm H_{0}=71~km~s^{-1}~Mpc^{-1}.$ The luminosity distance to the source is 2.4 Gpc, the image scale is 5.6~pc/mas \citep{Jorstad2017}.

\section{Observations and Data Analysis \label{sec:obs}}

4C~$+$21.35 was observed during a KaVA monitoring campaign of M87 \citep{Hada2017}. During the allocated epochs, observations were performed bi-weekly and quasi-simultaneously at 22 and 43~GHz, with a separation of one day between the two frequencies (when both were available). Observations at 22~GHz started in September 2014, followed by joint 22 and 43~GHz observations from May 2015 on.
Continuous bi-weekly observations began in 2016. In total we obtained thirteen epochs of data for 4C~$+$21.35 at 22~GHz and eleven at 43~GHz (see Table~\ref{tab:obs} for an overview). Angular resolutions (beam sizes) are on average 1.2$\times$1.1~mas and 0.8$\times$0.6~mas at 22 and 43~GHz, respectively. The on-source time is 70 minutes on average, the data are recorded at a rate of 1Gbps with 2-bit sampling: 16 IFs with a bandwidth of 16~MHz per IF were used in 2014, 8 IFs with a bandwidth of 32~MHz per IF were used from 2015 on. Only left-hand circular polarization was observed. In mid-November 2014, a $\gamma$-ray flare from 4C~$+$21.35 was detected by the \emph{Fermi}--LAT. Our observation closest in time was performed in January 2015. All data were correlated at the Korea-Japan Correlation Center (KJCC) at the Korea Astronomy and Space Science Institute (KASI) \citep{Oh2015,lee2015}.

We processed our KaVA data with the \texttt{AIPS} software package,\footnote{\url{http://www.aips.nrao.edu/index.shtml}} following the standard VLBI data reduction procedures. A-priori amplitude calibration was performed with the measured opacity-corrected system temperature and the gain curve of each antenna. Then, we ran the fringe-fitting process \texttt{FRING} to calibrate the visibility phases. We clipped out both the first and the last 50 channels (out of 1024 in total) to eliminate known band-edge artifacts. We constructed high resolution VLBI images with \texttt{DifMAP} \citep{Shepherd1994} using natural weighting for all epochs. Using the \texttt{modelfit} task, we fitted circular Gaussian model components to the brightest features in the ($u$,$v$) plane; this choice reduces the number of free parameters and provides consistent positions over multiple epochs \citep{MOJAVE6}. We removed components that overlapped with others or had a best-fit size of zero. We stopped model fitting when the residual map became consistent with the noise level of $\sim$1~mJy/beam.

The cross-identification of individual jet components from epoch to epoch is one of the key constraints on any kinematic analysis. In our case, thanks to the slow evolution of the jet structure with time and frequent observations, the cross-identification process was straightforward usually. We extracted the 2-dimensional position information of the jet components and checked that the variation of the position angles is less than 10${^\circ}$. We then tracked only the radial distances of the jet components from the core component, which we presumed to be stationary. 
We estimated flux density errors according to \citet{Fomalont1999}. We followed \citet{MOJAVE6} for estimating the errors on the best-fit positions of the Gaussian jet components:$\sim$10 percent of the component size convolved with the beam size. 

\begin{figure}
\centering
\includegraphics[trim=-2mm 5mm 0mm 20mm, clip, width=75mm]{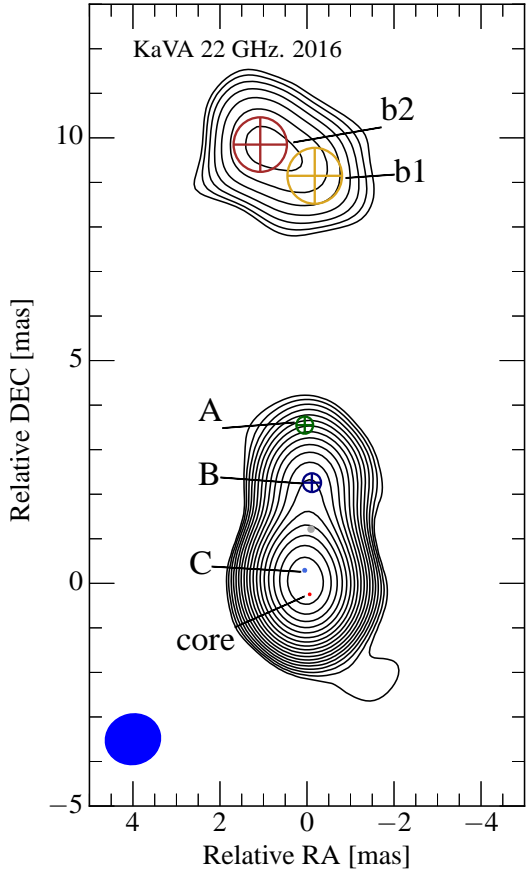}
\caption{A 22~GHz KaVA image of 4C~$+$21.35. The contour levels start at three times the rms noise value (0.64~mJy/beam), and increase in steps of $\sqrt{2}$. Circles labeled with letters mark identified jet components, the grey crossed circle is excluded from our analysis. The blue ellipse on the bottom left illustrates the CLEAN  beam (1.27 mas $\times$ 1.13 mas).  \label{fig:22Gcon1}}
\end{figure}

\begin{figure}
\centering
\includegraphics[trim=5mm 0mm 0mm 0mm, clip, width=85mm]{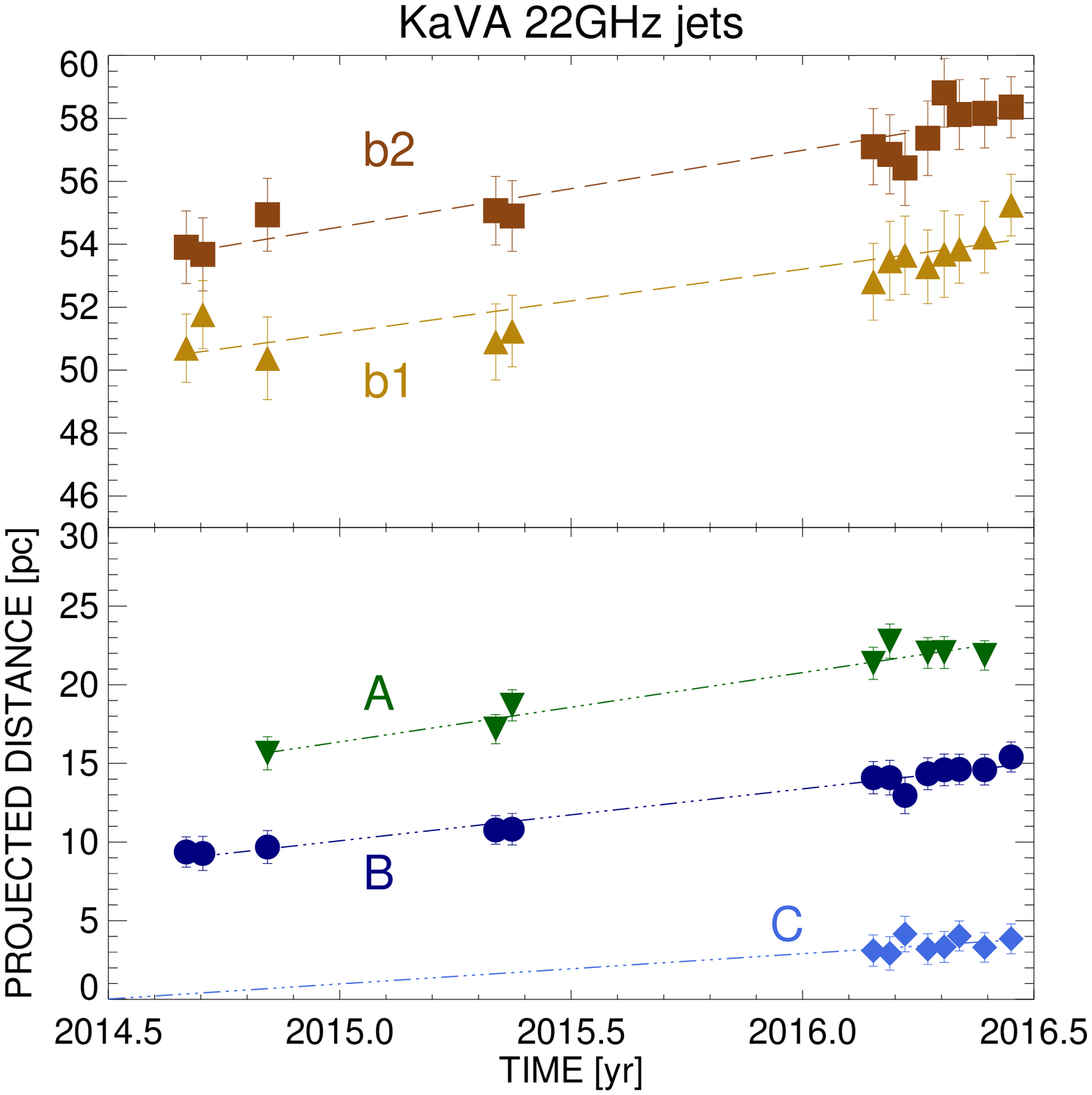} \\[3mm] 
\includegraphics[trim=6mm -4mm 6mm 0mm, clip, width=84mm]{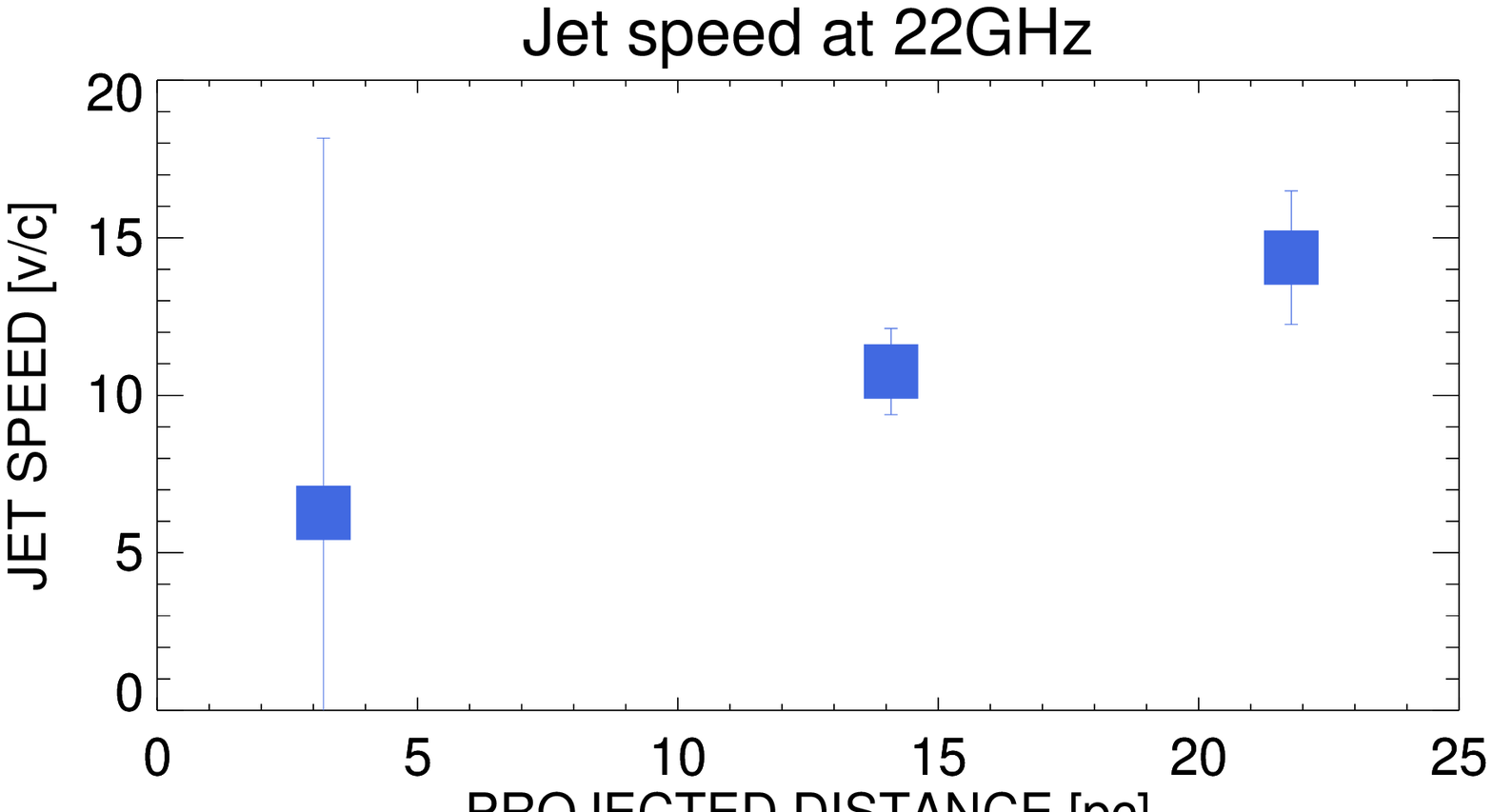} \\[1mm]
\caption{Jet kinematics derived from the 22~GHz KaVA maps. \emph{Top panel:} Projected distance from the core as a function of time for all five jet components, separately for the outer components in the 10-mas blob (upper diagram) and the three inner jet components (lower diagram). \emph{Bottom panel:} Apparent speeds of the three inner components as a function of projected distance from the core. \label{fig:22G_KaVA}}
\end{figure}

\begin{figure}
\centering
\includegraphics[trim=0mm 25mm 0mm 35mm, clip, width=75mm]{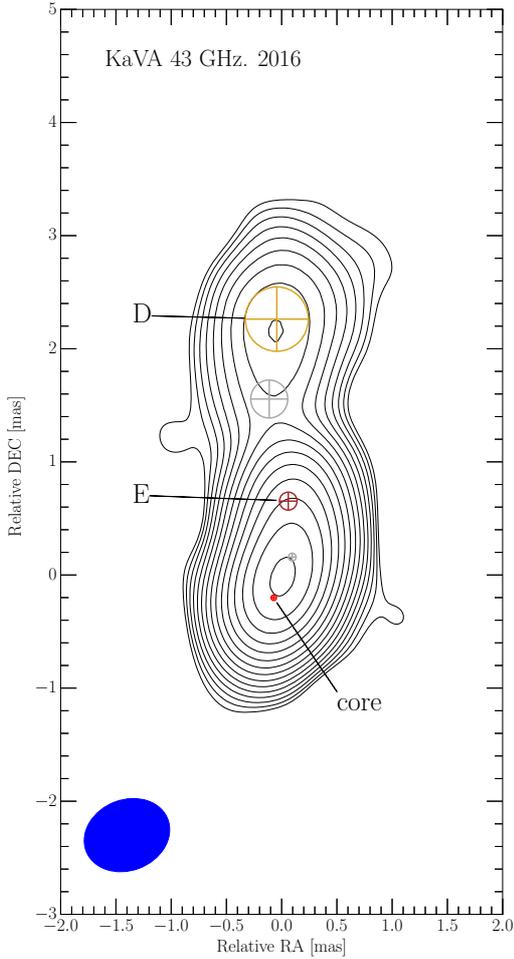}
\caption{A 43~GHz KaVA image of 4C~$+$21.35. The contour levels start at three times the rms noise value (1.14~mJy/beam), and increase in steps of $\sqrt{2}$. The two circles labeled with letters mark jet components that we were able to track reliably, the two grey crossed circles mark jet components we could not re-identity or track reliably. The CLEAN beam size is illustrated on the bottom left (0.79 mas $\times$ 0.63 mas). \label{fig:43Gcon1}}
\end{figure}

\begin{figure}
\centering
\includegraphics[trim=5mm -2mm 0mm 0mm, clip, width=85mm]{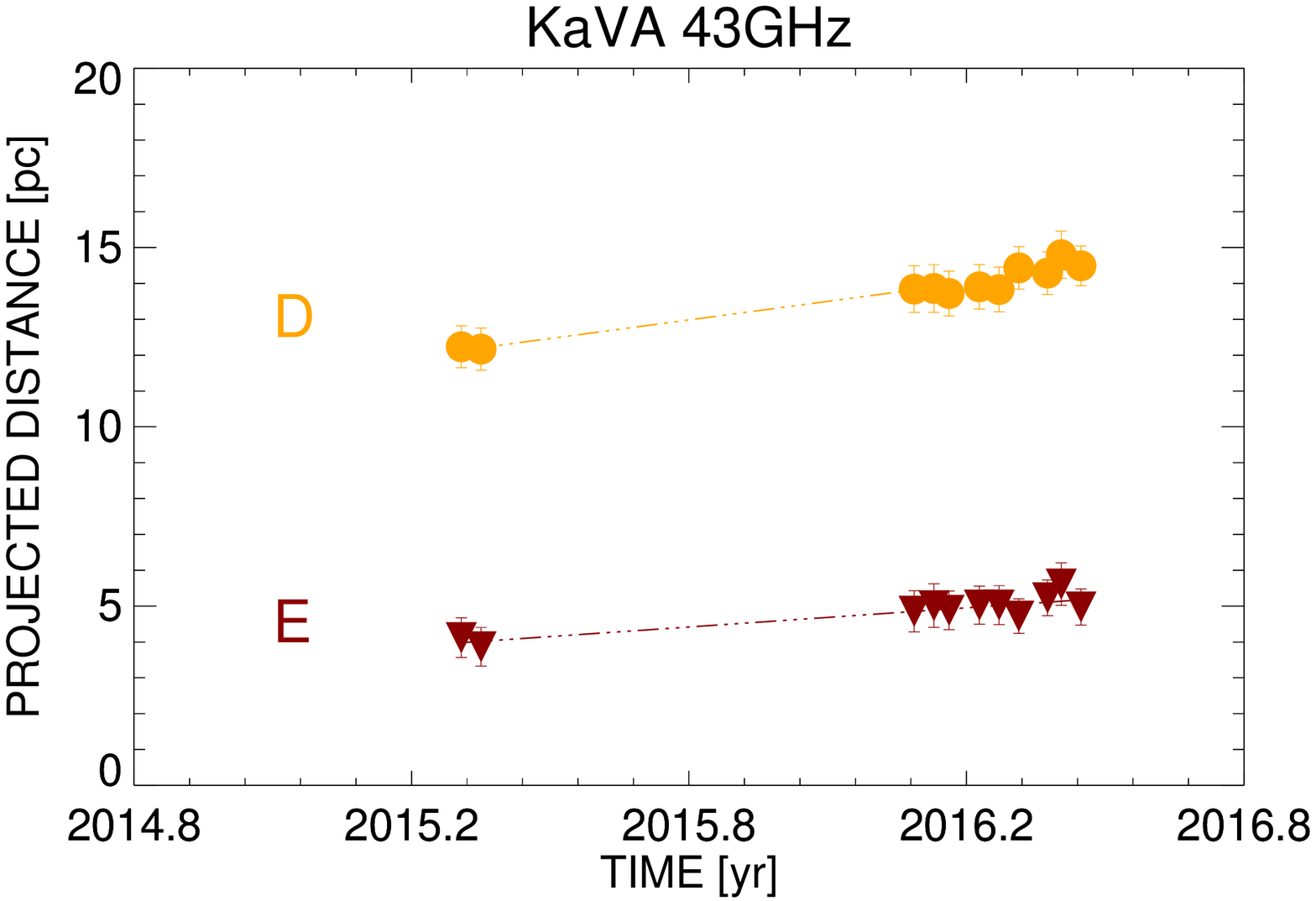} \\[2mm] 
\includegraphics[trim=5mm -4mm 0mm 0mm, clip, width=85mm]{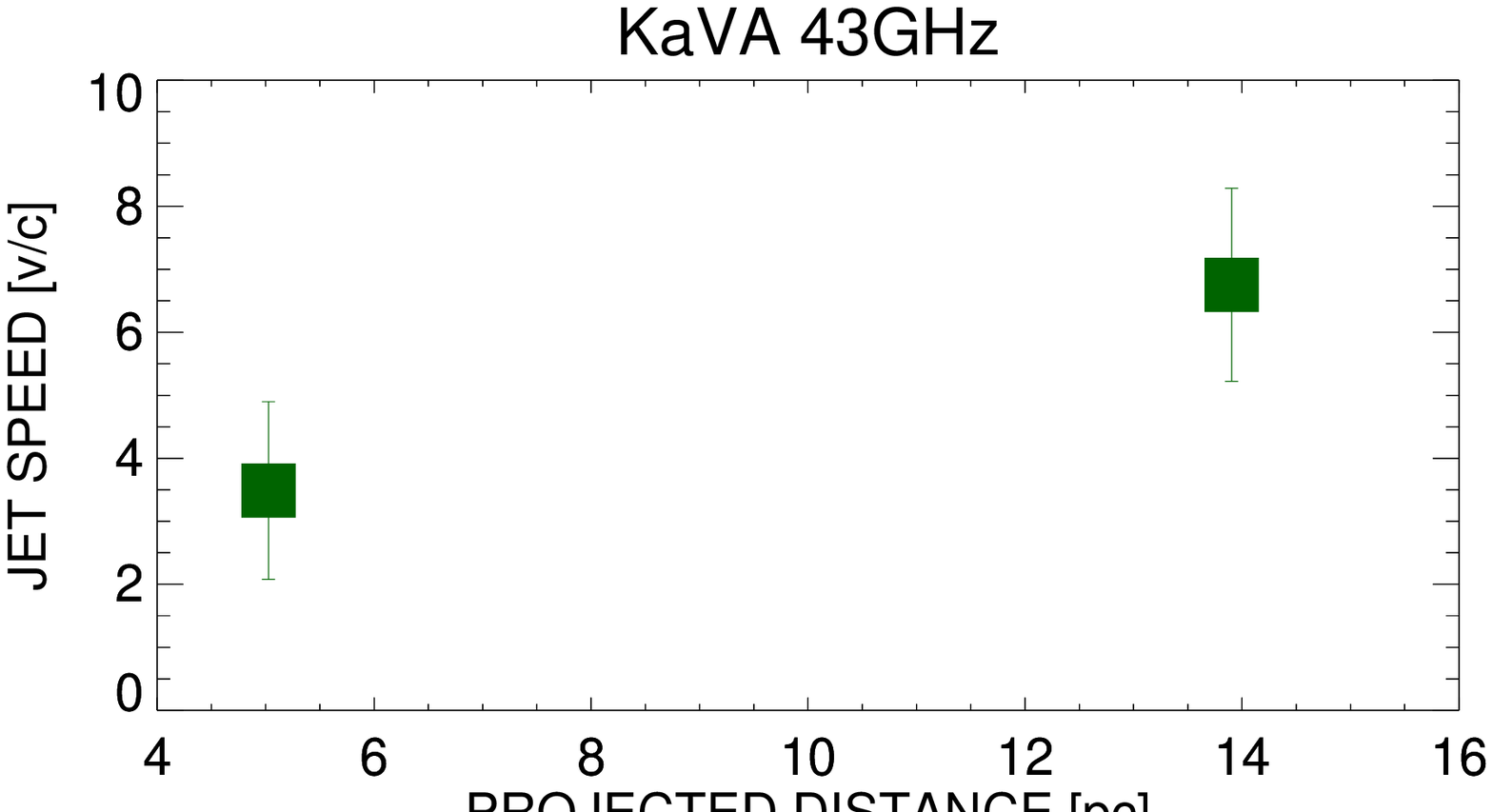}
\caption{Jet kinematics derived from the 43~GHz KaVA maps. \emph{Upper panel:} Projected distance from the core as a function of time for the two components we could track reliably. \emph{Lower panel:} Apparent jet speed as a function of projected distance from the core for the two components shown in the upper panel.  \label{fig:43G_KaVA}}
\end{figure}

\begin{figure}
\centering
\includegraphics[width=85mm]{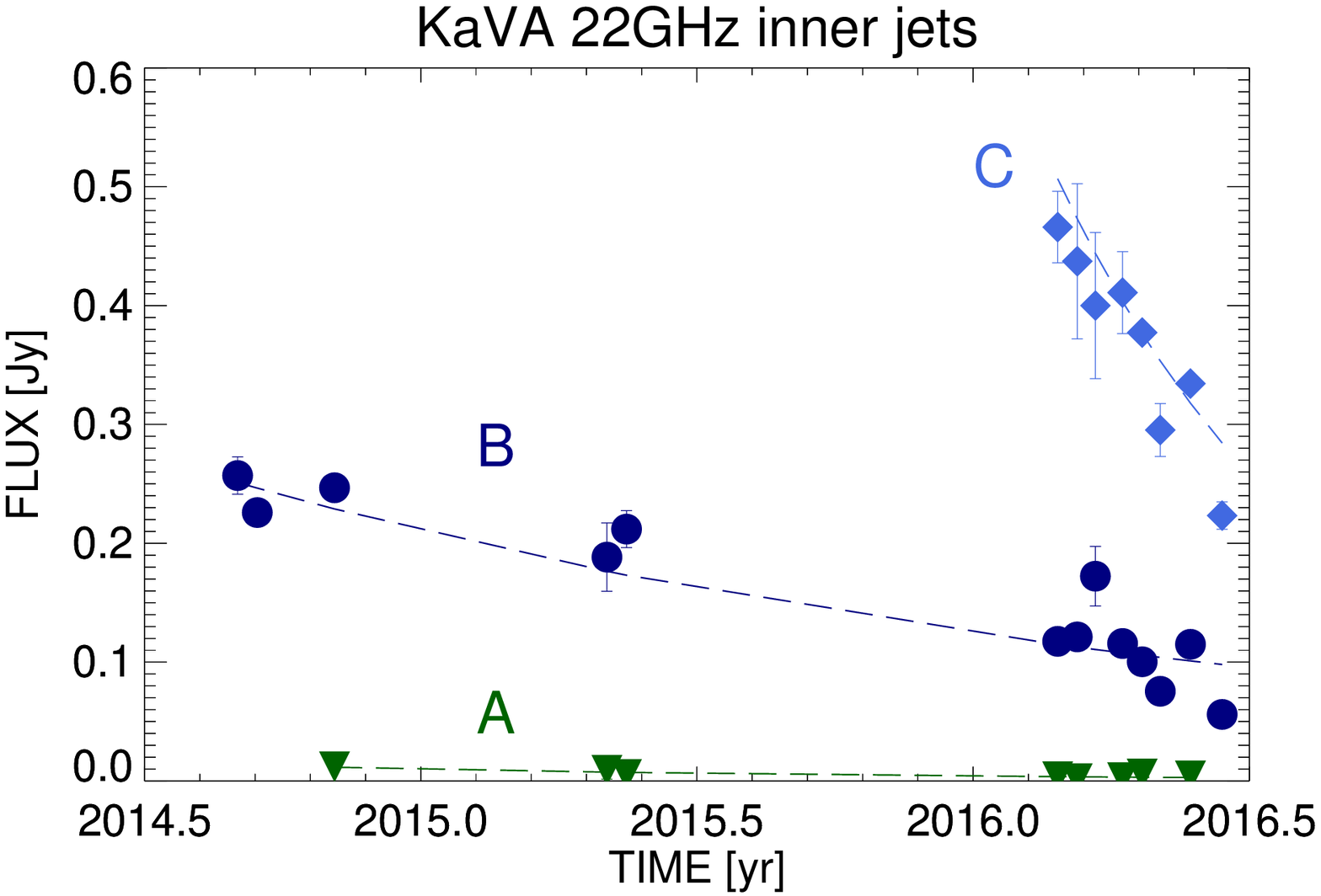} \\ 
\includegraphics[width=85mm]{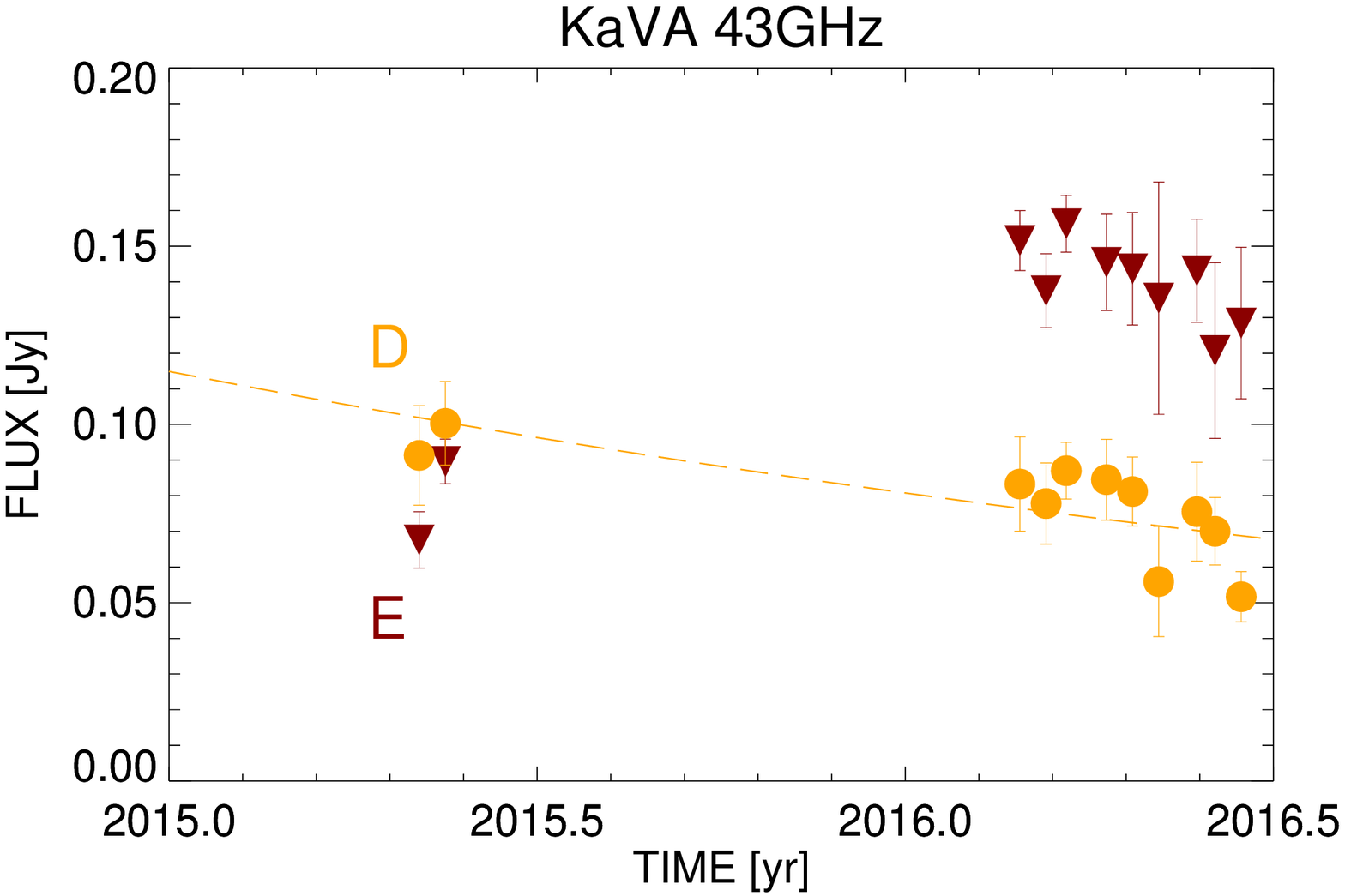}
\caption{Flux evolution of KaVA jet components. \emph{Top panel:} For the three inner jet components at 22~GHz. \emph{Bottom panel:} For the two components at 43~GHz. \label{fig:FluxDecay}}
\end{figure}

\section{Results \label{sec:results}}
\subsection{KaVA at 22 GHz \label{sec:22GHz}}

In our 22~GHz maps, 4C~$+$21.35 has a compact straight jet extending up to 4~mas to the north (cf. Figure~\ref{fig:22Gcon1}). In addition, we detect an extended structure (``blob'') located about 10~mas north of the core, at a position angle of about $10^{\circ}$ (counter-clockwise from north).

We identified three components in the inner jet and two in the blob. The middle component in the inner jet was excluded from our analysis -- even though we detected it throughout the observations -- because it showed significant, random position variations; what we observed is probably a blend of multiple components. The innermost component became visible in early 2016; accordingly, we suspect that it was blended with the radio core until then. This is supported by the fact that the peak flux density of the core decreased from 1.6 to 0.9~Jy from 2015 to 2016. All the jet components show linear outward motion; the further from the core, the faster they move (see Figure~\ref{fig:22G_KaVA}). The apparent speeds are $(6.3\substack{+11.9\\-6.3})c$, $(10.8\pm1.4)c$, and $(14.4\pm2.1)c$ outward: For the newly ejected jet component, we could constrain only the upper limit of speed. Extrapolating the position of the newly emerged component (component C in Figure~\ref{fig:22Gcon1})  back to zero distance from the core suggests an ejection in $2014.5\pm3.4$, assuming a constant component speed, consistent with the peak time of the mid-November 2014 $\gamma$-ray flare.

The 10-mas blob moved toward the north-east with an apparent speed of $(5.7\pm0.4)c$. Its flux density did not change significantly during our observations. The trajectories of the two jet components inside the blob are almost perpendicular to the jet orientation. This is consistent with previous MOJAVE observation results \citep{MOJAVE7}.

\subsection{KaVA at 43 GHz \label{sec:43GHz}}

At 43~GHz, only the core and the inner jet of 4C~$+$21.35 are visible; the 10-mas blob is too faint to be seen at this frequency (cf. Figure~\ref{fig:43Gcon1}). The core peak flux dropped from 1 Jy to 0.5 Jy from 2015 to 2016, consistent with the observations at 22~GHz. We detected four jet components and could track two of them. The component closest to the core, at a distance of around 0.3~mas, seemed to be blended with the core. The component at around 1.7~mas showed no clear preferred direction of movement. We were not able to confirm the new knot whose appearance in 2015 \citet{Troitskiy2016} reported, or the stationary component located at 0.14~mas from the 43~GHz core. This is likely due to limited resolution since our maps show that the innermost component is located at $\sim$0.3~mas and exhibits random movements. Alternatively, it could have been short-lived and was not detectable when our observation re-started in the beginning of 2016.

The jet components moved outward monotonously throughout our observations (see Figure~\ref{fig:43G_KaVA}). The flux density from the outer component decreased almost linearly, while the flux density from the inner component \emph{increased} (cf. Figure~\ref{fig:FluxDecay}). The apparent jet speeds are $(3.5\pm1.4)c$ and $(6.8\pm1.5)c$, respectively.

\subsection{Intrinsic Jet Speeds and Doppler factors \label{sec:jetspeeds}}

The apparent (plane of sky) jet speed in units of speed of light $\beta_a$, intrinsic jet speed $\beta$, and viewing angle between the jet axis and the line of sight $\theta$ are related via 
\begin{equation}
\beta = \frac{\beta_{a}}{\sin\theta+\beta_{a} \cos\theta} ~ .
\end{equation}
As we observe apparent jet speeds up to $13c$, we can infer that $\theta \lesssim 10^{\circ}$, in agreement with the value $\theta\approx5.3^{\circ}$ reported by \citet{MOJAVE14}. Using the canonical value ($\theta=5.3^{\circ}$), we can derive intrinsic jet speeds of $(0.990+0.010)c$ to $(0.998\pm0.001)c$ at 22~GHz and $(0.978\substack{+0.007\\-0.017})c$ to $(0.991\substack{+0.002\\-0.004})c$ at 43~GHz. Consequently, the Doppler factors we obtain are  9.8, 10.2, and 10.8 at 22~GHz and 8.0 and 10.0 at 43~GHz. These values are consistent with the earlier findings by \citet{MOJAVE14}.

\subsection{Flux Evolution and Synchrotron Cooling \label{sec:cool}}

Most of the jet components we observed with KaVA showed a steady decrease in flux density with time. Especially the three innermost 22~GHz jet components showed well-sampled decay light curves (cf. Figure~\ref{fig:FluxDecay}). We described those with an exponential decay model
\begin{equation}
F = X \exp(-Y t)  + Z
\end{equation}
where $t$ is the time, $X$, $Y$, and $Z$ are constants, and $Y=1/\tau$ with $\tau$ being the decay time scale. The best-fit decay time scales for each jet component are for A: $1.1 \pm 0.2$ yr, B: $1.9 \pm 0.1$ yr, C: $0.5 \pm 0.04$ yr at 22 GHz, and for D: $2.8 \pm 0.8$ yr at 43 GHz, respectively. Assuming that the flux decay results from synchrotron cooling, we may equate the decay time scale $\tau$ with the synchrotron cooling time scale $\tau_{\rm cool}$.
For a relativistic plasma, where the electron energy distribution follows a broken power law, and considering the dominant synchrotron energy is emitted near the break frequency in the electron spectrum, 
we can estimate the electron Lorentz factor $\gamma$ from the ratio of the peak frequencies of inverse Compton ($\nu_{\rm IC}$) and synchrotron ($\nu_{\rm syn}$) emission like \citep{Beckmann2012}
\begin{equation}
\gamma = \sqrt{\frac{3 \nu_{\rm IC}}{4 \nu_{\rm syn}}} ~ .
\end{equation}
For 4C~$+$21.35, $\nu_{\rm IC} \sim 10^{24}$ Hz and $\nu_{\rm syn} \sim 10^{16}$ GHz (from Figure~9 of \citealt{Ackermann2014}), implying $\gamma \sim 9\,000$. For a cooling synchrotron plasma, we can connect the cooling time scale to $\gamma$ and the magnetic field strength $B$ via \citep{rybicki1979}
\begin{equation}
\tau_{\rm cool} = 7.74 \left[\frac{\delta}{1+z}\right]^{-1} B^{-2} \gamma^{-1} ~~~ \rm seconds .
\label{eq:tau}
\end{equation}
For a redshift $z=0.43$ and a Doppler factor $\delta \sim 10 $, cooling times in the range 0.5--2.8 years imply $B \approx 1-3~\mu$T.

\begin{table}
\centering
\caption{Best-fit parameters for each Gaussian jet component, for the period April 21--22, 2016. \label{tab:modelfit}}
\begin{tabular}{cccccccccc}
\hline
Frequency & ID & Flux & Distance & Position angle & Major axis \\
(GHz) & & [Jy] & [mas] & [${\circ}$]  & [mas] \\
\hline

22 & A  & 0.0058 & 3.7 & 4.6 & 0.43 \\ 
	 & B  & 0.10      & 2.3 & -2.3& 0.38 \\ 
	 & C  & 0.38     & 0.33 & 13 & 0.023 \\\\  
	 & b1 & 0.033   & 9.3  & 2.3 & 1.7  \\ 
	 & b2 & 0.021   & 10  & 7.9  &  0.84 \\\\ 

43 &	D &	0.081 & 2.3  & -1.5  & 0.58 \\
	 &	E &	0.14   &	 0.72 & 5.7  & 0.16  \\

\hline
\end{tabular}
\end{table}

\subsection{Spectral Index Maps \label{sec:specmap}}

\begin{figure}
\centering
\includegraphics[trim=2mm -2mm 10mm 3mm, clip, width=80mm]{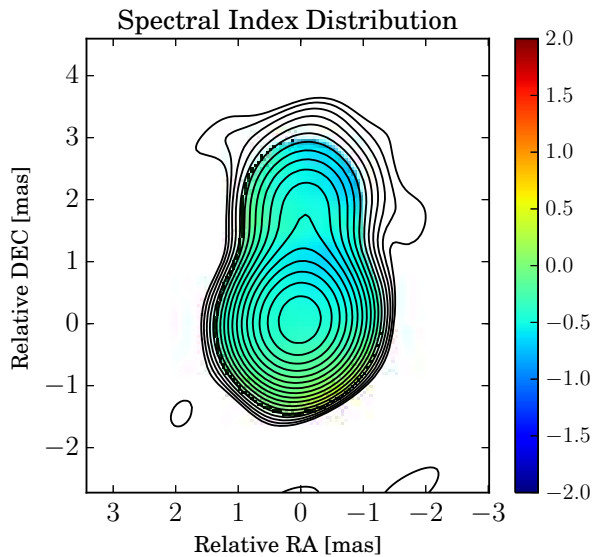} 
\caption{KaVA spectral index map of 4C~$+$21.35 from May 2015 data.  \label{fig:spmap1}}
\end{figure}

Since the 22~GHz and 43~GHz observations were performed quasi-simultaneously with one day separation between the frequency bands, spectral information is available. Our spectral index $\alpha$ is defined via $S_{\nu}\propto\nu^{\alpha}$. We obtained seven spectral index maps of the source using the \texttt{VIMAP} software routine \citep{Kim2014}: two in 2015 and five in 2016. However, 4C~$+$21.35 is compact and core-dominated. Accurate alignment of maps obtained at two different frequencies requires alignment of the (presumably) optically thin jet at 22 and 43~GHz via spatial cross-correlation; in our case, only small parts of the source could be used. Even so, our spectral index maps managed to capture the smooth transition from a flat spectrum ($\alpha\approx0$) at the core to a steep spectrum ($\alpha\approx-1$) at the very tip of the jet region, indicating that the core is optically thick while the jet is optically thin -- as it was to be expected. Spectral index maps obtained at different epochs do not show significant differences; accordingly, we only present one of them in Figure~\ref{fig:spmap1} as a representative example.


\section{Discussion \label{sec:discuss}}
\subsection{Boston University Program Data \label{sec:BUdata}}

In order to cross-check our 43~GHz KaVA results with VLBA data, we used archival data from the Boston University Blazar Monitoring Program (BU Program).\footnote{\url{www.bu.edu/blazars/VLBAproject.html}} The BU Program archive provided us with eight epochs of 43~GHz maps obtained from April 2015 to April 2016. During monitoring campaigns, the BU Program gathers data on a monthly basis. VLBA has more antennas and a better $uv$-coverage than KaVA. As a result, we found more circular Gaussian jet components in the (original) VLBA images than in our KaVA maps. For a direct comparison to our data, we step by step trimmed out components from the BU Program maps: we removed peripheral components and repeated the component fitting procedure until we came up with jet component locations and sizes that are as close as possible to the KaVA data. We then continued with an analysis of the jet kinematics as we did for the KaVA data.

\begin{figure}
\centering
\includegraphics[trim=6mm 0mm 2mm 0mm, clip, width=85mm]{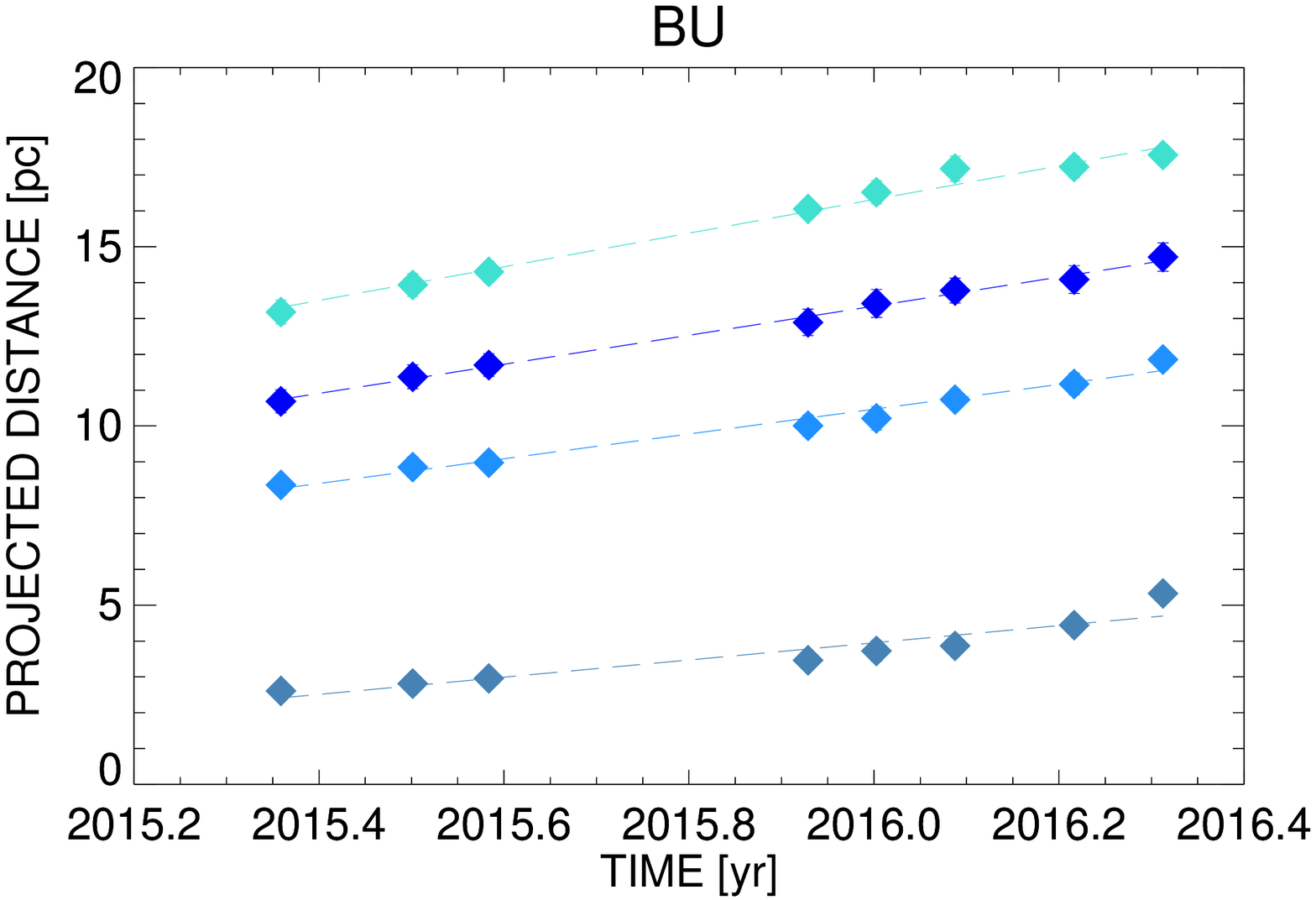} \\[2mm]   
\includegraphics[trim=6mm -4mm 4mm 0mm, clip, width=85mm]{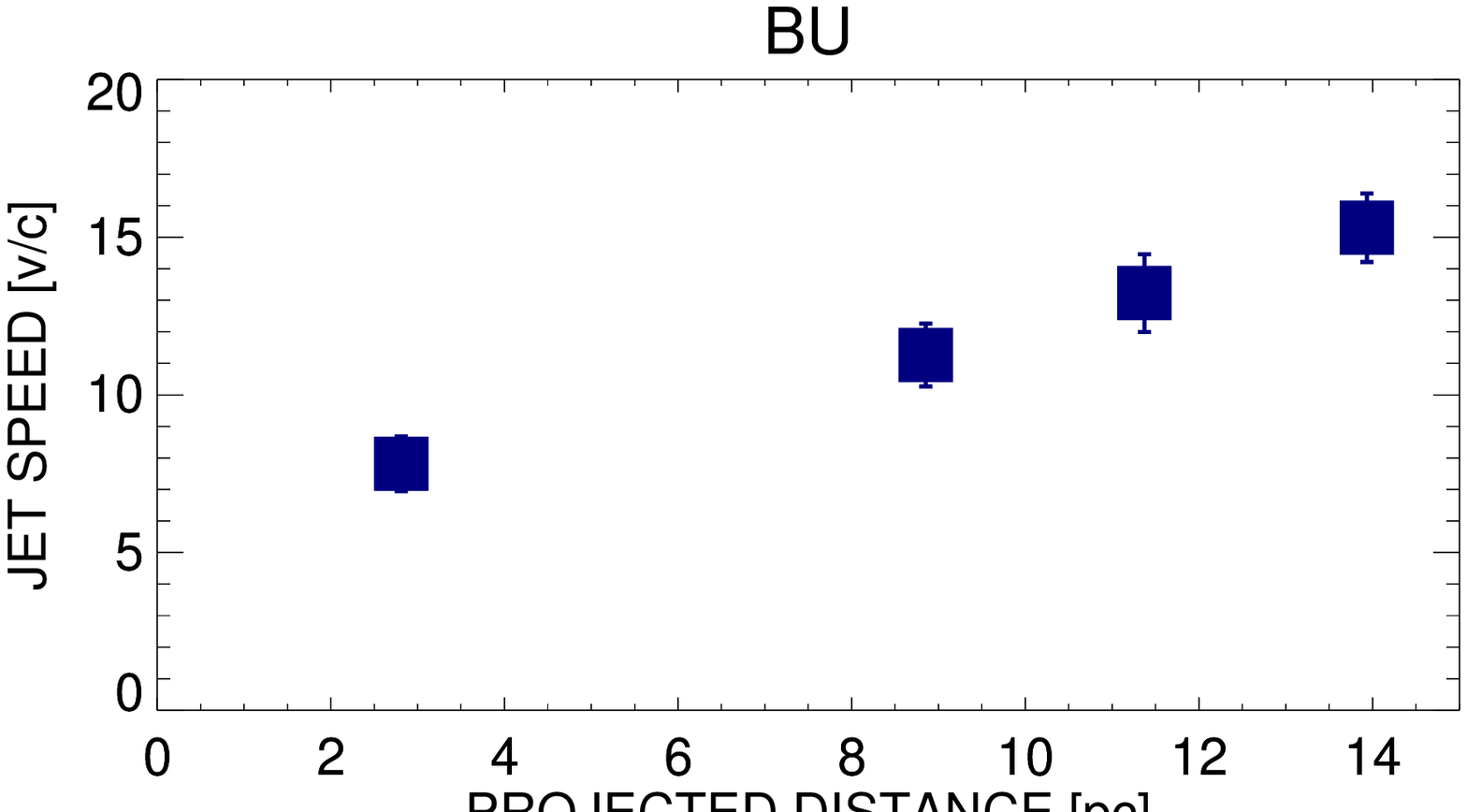}
\caption{Jet kinematics derived from BU Program 43~GHz data. \emph{Top panel:} Core distance as a function of time for four jet components. \emph{Bottom panel:} Jet component speed as a function of core distance. Error bars represent $1\sigma$ uncertainties. \label{fig:BU}}
\end{figure}

We found apparent jet speeds ranging from $(7.8\pm0.9)c$ to $(15.3\pm1.1)c$, with higher speeds at larger distances from the core (see Figure~\ref{fig:BU}). These speeds are faster than the ones we obtained from our 43~GHz KaVA data by a factor of at least two. Since the observing frequencies are identical and the observations overlap in time, we suspected blending effects due to differences in angular resolution as a main cause for this discrepancy; the beam-size of VLBA at 43~GHz is 0.38 $\times$ 0.19 mas, half of the KaVA beam-size.

\subsection{BU Program Data with KaVA {\itshape\bfseries uv}-coverage \label{sec:clippedBUdata}}

\begin{figure}
\centering
\includegraphics[trim=3mm 0mm 4mm 1mm, clip, width=85mm]{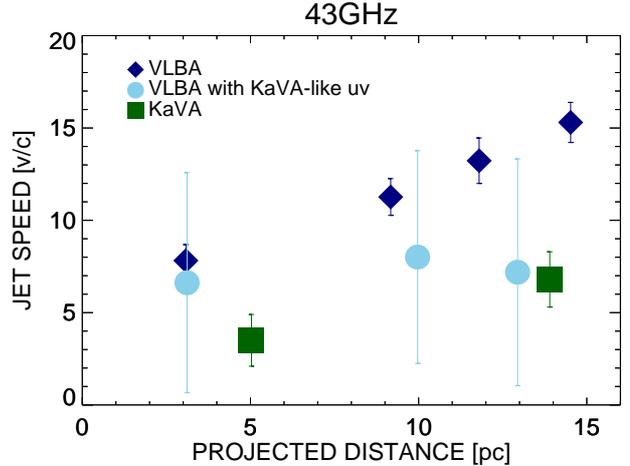}
\caption{Jet component speeds as a function of core distance for three different 43~GHz datasets. Blue: VLBA; green: KaVA data; skyblue: VLBA with KaVA-like $uv$ coverage. Error bars represent $1\sigma$ uncertainties. \label{fig:JetAccBUk}}
\end{figure}

In order to make the $uv$-coverage of the BU Program data as similar as possible to our KaVA observations, we removed visibilities obtained at baselines longer than those of KaVA, i.e., $3 \times 10^8 \lambda$ ($\lambda$ being the wavelength). The resulting $uv$-coverage still shows a higher density of visibilities than the KaVA case. The beam size is $0.58 \times 0.53$ mas (July 2015 data) and the image noise rms value is $\sim$1~mJy/beam, both of which are comparable to that of KaVA. In the modified images, we again identified and tracked the jet components if possible. The jet component locations are consistent with those in the KaVA maps. For the components at 3~pc and 10~pc from the core, the apparent speeds are comparable to the original BU data; at 13~pc, however, the estimated speed was $\sim7c$. This is just about half of the speed found in the original VLBA data and close to the value for the KaVA component at 14~pc (see Figure~\ref{fig:JetAccBUk}).

\subsection{MOJAVE Data \label{sec:MOJAVEdata}}

As a further cross-check, we analyzed archival 15.4~GHz VLBA data obtained by the MOJAVE program\footnote{\url{www.physics.purdue.edu/MOJAVE/allsources.html}} and compared those to our 22~GHz KaVA data. The MOJAVE archive provided us with seven epochs of data spanning from September 2014 to September 2016.

\begin{figure}
\centering
\includegraphics[trim=6mm 0mm 0mm 0mm, clip, width=85mm]{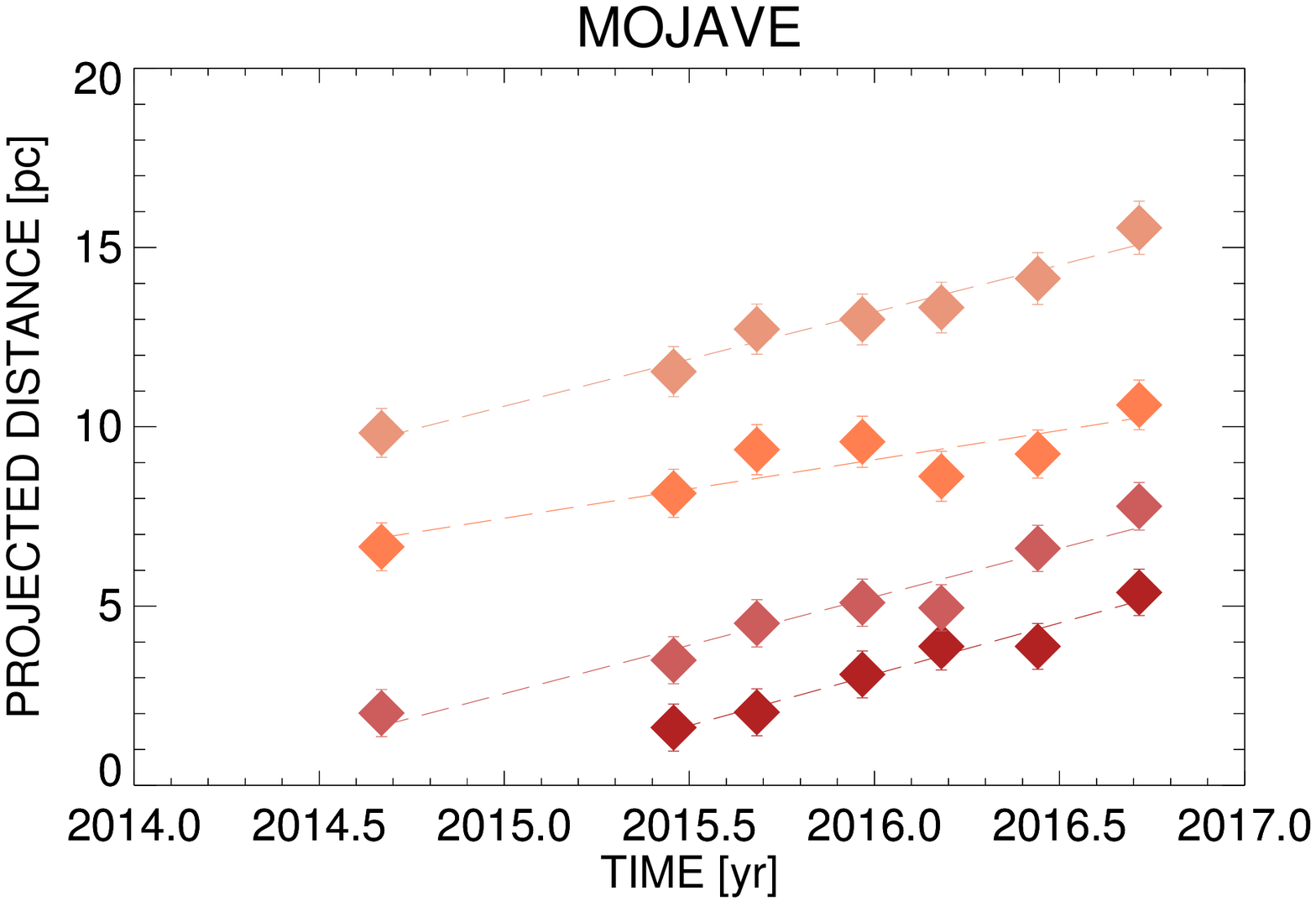} \\[2mm]   
\includegraphics[trim=6mm -4mm 0mm 0mm, clip, width=85mm]{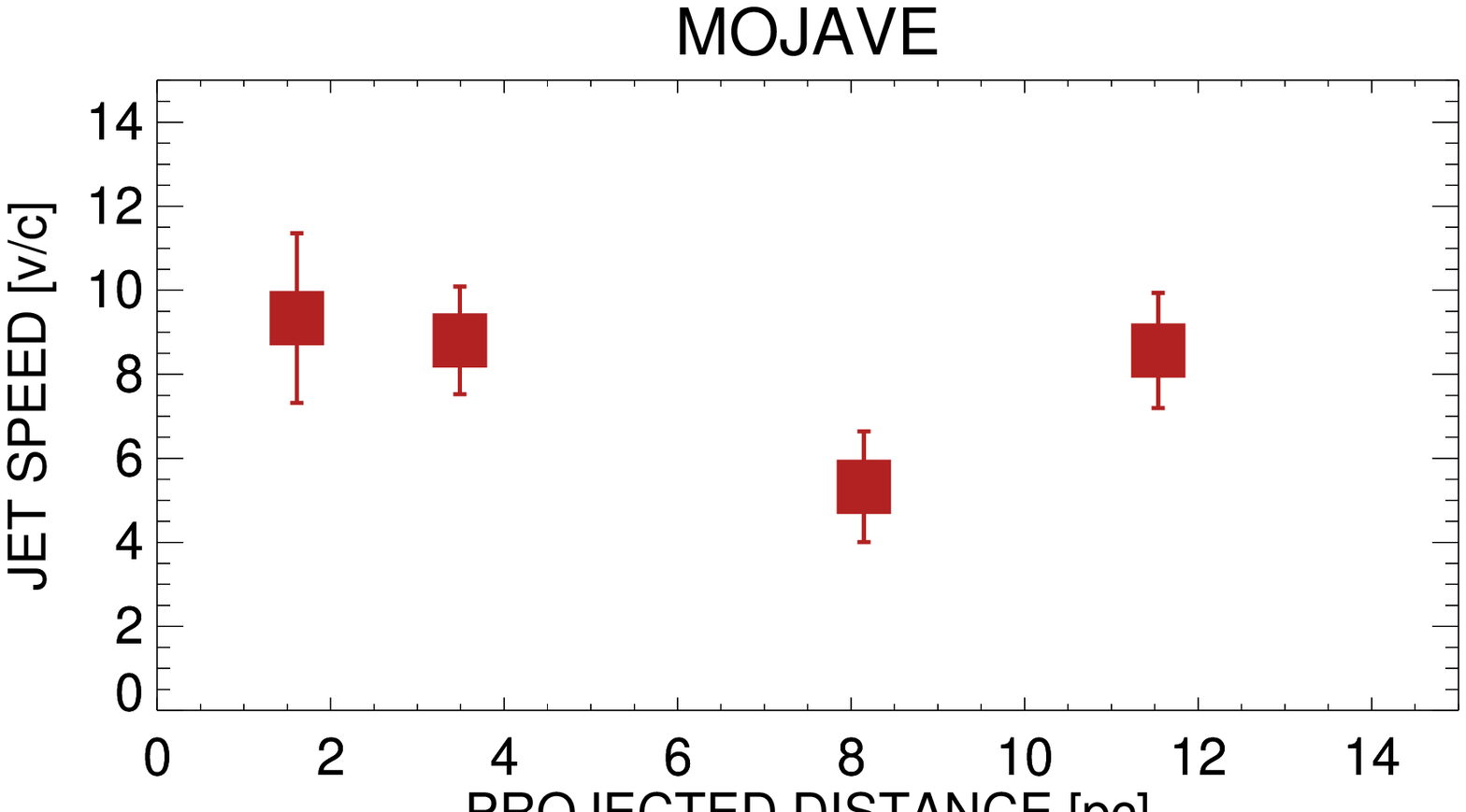}
\caption{Jet kinematics derived from MOJAVE 15.4~GHz data. \emph{Top panel:} Core distance as a function of time for four jet components. \emph{Bottom panel:} Jet component speed as a function of core distance. The $1\sigma$ error bars are smaller than the symbols. \label{fig:MOJAVE}}
\end{figure}

The beam size of the MOJAVE maps, $1.01 \times 0.576$ mas (June 2015 data) and the map rms value, 2~mJy/beam, are about the same as that for KaVA at 22~GHz. We thus proceeded with direct comparison of the two datasets without modifying the VLBA maps. We found four jet components at 1, 3, 8, and 12~pc from the core, closer to the core than those found in the KaVA 22~GHz data. The apparent jet component speeds range from $(5.3\pm1.3)c$ to $(9.3\pm2.0)c$, comparable to the KaVA 22~GHz data (see Figure~\ref{fig:MOJAVE}).

\subsection{The 10-mas-blob \label{sec:10masblob}}

The 10-mas-blob continues to drift to the north-east, in agreement with the report by \citet{MOJAVE7}. This is interesting for two reasons: (a) the 10-mas-blob has remained on a stable bent path since its ejection in the early 2000s; (b) the trajectory of the 10-mas-blob, when extrapolated, is inconsistent with that of an older blob located $\sim$15 mas (not shown in our maps) that shows an almost radial trajectory. Non-ballistic motion, as observed for the 10-mas-blob, is a well-known feature of blazar jets; \citet{MOJAVE6} and \citet{Kellermann2004} found almost one third of the blazars in their samples to show non-radial motion. \citet{MOJAVE7} concluded that accelerations perpendicular to the jet direction are common in blazars. They also found that, despite changes in jet direction, jet components have an overall tendency to follow the downstream flow; this was also noted by \citet{Kellermann2004} who concluded that jet flows follow pre-existing bent channels. Even so, the mechanism behind non-radial motions is not yet understood. Given that many blazars show misaligned jet orientations on parsec to kiloparsec scales \citep{Kharb2010}, it is possible that the 10-mas blob changes its course again in the future. Based on Very Large Array 1.4~GHz maps, \citet{Cooper2007} reported that the blob seems to follow the existing large-scale structure to the east.

\section{Conclusions \label{sec:conclusion}}

We have been able to constrain the kinematics of the jet of 4C~$+$21.35 with KaVA observations spanning from mid-2014 to mid-2016. Our results demonstrate the ability of KaVA to provide important information on AGN jet kinematics in the frame of focused observing campaigns, in line with several recent works \citep{Niinuma2014, Oh2015, Hada2017}. Exploiting the dense (bi-weekly) sampling of our data, we were able to identify and track jet components from epoch to epoch with high reliability. Due to the timing of our observations, we were able to find a connection between the $\gamma$-ray flare of 2014 and the jet kinematics. Furthermore, we were able to accurately trace the exponential flux decay for the inner jet components, especially for the newly ejected jet component C at 22~GHz.

Our maps resolve several components within the northward-pointing jet. At 22~GHz (Section~\ref{sec:22GHz}), we identify five components in total, three of them in the continuous inner jet within 4~mas from the core. Notably, a new component emerged in early 2016. This new component moved along the straight path of the other two inner jet components; the apparent speeds of the three components are $(6.3\substack{+11.9\\-6.3})c$, $(10.8\pm1.4)c$, and $(14.4\pm2.1)c$ (outward from the core) and are in good agreement with those obtained from MOJAVE 15.4~GHz data (about $8c$).


At 43~GHz (Section~\ref{sec:43GHz}), we have been able to resolve the (inner) jet into four components out which we could re-identify (from epoch to epoch) and track two reliably. Even though all the jet components appear to move straight to the north, the jet appears somewhat bent (with a tilt of about $20\deg$ to the west) within the innermost $\lesssim$0.5~mas. Indeed, the innermost jet component, located about 0.3~mas away from the core, seems to mark the approximate location where the jet flow turns straight north (see Figure~\ref{fig:43Gcon1}).

The apparent KaVA jet component speeds at 43~GHz are $(3.5\pm1.4)c$ and $(6.8\pm1.5)c$ at 5 and 14 pc, respectively. However, a kinematic analysis of the contemporaneous BU Program 43~GHz (Section~\ref{sec:BUdata}) data finds speeds ranging from $(7.8\pm0.9)c$ to $(13.2\pm1.2)c$. Likewise, \citet{Troitskiy2016} reported speeds of $9c$ to $22c$ for five jet components located within $\lesssim$5 pc using BU Program data obtained from 2010 to 2016. We suspect an effect of angular resolution: the beam-size of the BU Program data, 0.38~mas~$\times$~0.19~mas, is about half of that of KaVA at 43~GHz, 0.72~mas~$\times$~0.63~mas. Our test analysis of BU Program data with KaVA-like $uv$-coverage (Section~\ref{sec:clippedBUdata}) supports this conclusion: it found a substantial drop of the apparent jet speed compared to the one derived from the original VLBA maps; this ``loss of speed'' increases with increasing core distance and reaches up to about 50\%. The ``loss of speed'' effect is, arguably, a consequence of the (well known) fact that, physically, an AGN jet is not a group of discrete point-like sources but a complex continuous distribution of matter. Observations with different $uv$-coverages will beam-average different regions of a jet which may lead to systematic differences in component positions as function of time. Indeed, it seems somewhat unsettling that the observed kinematics may depend on the instrument used for observations.

Another potential mechanism leading to a ``loss of speed'' effect is \emph{time} resolution: observations obtained at discrete epochs may not reliably re-identify jet components from epoch to epoch. \citet{Piner2007} noted that temporal undersampling often results in misidentification of components and under- or overestimation of apparent jet speeds. Notably, the high cadence of KaVA observations -- every two weeks since early 2016 -- reduces the risk of time resolution effects distorting kinematic analyses. In the case of 4C~$+$21.35, proper motions of jet components are below 1~mas~yr$^{-1}$, corresponding to about 40~$\mu$as per two weeks and thus making a confusion of components is unlikely.


In Section~\ref{sec:cool}, we found $1/e$ flux decay time scales for the KaVA jet components ranging from $\sim0.5$~years to $\sim3$~years. Equating this time scale with the synchrotron cooling time provided us with the estimated magnetic field strengths of about 1 to 3~$\mu$T. We used an electron Lorentz factor $\gamma\sim9\,000$, a value located at the upper end of the interval covered by values typical for FSRQ \citep{Ghisellini2002}. The magnetic field strengths we find are consistent with the values typical for blazars which are in the range from 1 to 10~$\mu$T \citep{Lewis2016}. \citet{MOJAVE9} independently calculated the magnetic field strengths at the core and at 1~pc downstream from the jet vertex using the core shift effect, finding 4~$\mu$T and 90~$\mu$T, respectively -- which is consistent with our findings.

We detected a new 22~GHz jet component in early 2016. Assuming a constant speed and extrapolating its motion back to zero distance from the core suggests that it was ejected at the end of October 2014. Although this estimated ejection time has relatively large uncertainties, it is in agreement with the time of the mid-November 2014 Fermi-LAT $\gamma$-ray flare. This is consistent with the very high energy emitting region being located close to the broad line region (thus avoiding absorption by BLR clouds) as well as a small size of the VHE emitter (as already expected from $\gamma$-ray / VHE variability) \citep{Kushwaha2014, Tavecchio2011}. \citet{Aleksic2014} also came to the same conclusion from the MAGIC observation of a FSRQ, PKS 1510-089. Likewise, \citet{Ackermann2014} constructed a SED with a one-zone model that assumes a very compact emission region outside the BLR.

Overall, our KaVA observations of 4C~$+$21.35 strengthen the case for a causal connection between the occurrence of $\gamma$-ray flares and the ejection of new jet components in blazars. This merits additional radio-interferometric studies of VHE AGN in order to probe this suspected connection in depth. Our analysis, however, clearly shows the importance of high cadence observations in order to avoid misunderstanding the jet kinematics.

\section*{Acknowledgements}
We are grateful to all staff members of KaVA who helped to operate the telescopes and process the data. KVN is a facility operated by the Korea Astronomy and Space Science Institute. VERA is a facility operated by the National Astronomical Observatory of Japan in collaboration with associated universities in Japan. The research has made use of the data from Boston University Blazar Monitoring Program (VLBA-BU-BLAZAR; \url{http://www.bu.edu/blazars/VLBAproject.html}), funded by NASA through the Fermi Guest Investigator Program. This research has made use of data from the MOJAVE database which is maintained by the MOJAVE team \citep{MOJAVE6}. The VLBA is operated by the National Radio Astronomy Observatory which is a facility of the National Science Foundation operated under cooperative agreement by Associated Universities, Inc. We acknowledge financial support from the National Research Foundation of Korea (NRF) via Basic Research Grant 2015R1D1A1A01056807. TA acknowledges the grant of the Youth Innovation Promotion Association of CAS. J. W. Lee is grateful for the support of the National Research Council of Science and Technology, Korea (Project Number EU-16-001). This work is partially supported by JSPS KAKENHI Grant Numbers JP18K03656 and JP18H03721.



\appendix

\section{Gaussian modelfit parameters for each jet component}

\begin{table}
\centering
\caption{Gaussian modelfit parameters for each jet component. Observation dates are indicated in bold face, quasi-simultaneous observations are grouped together. \label{tab:modelfit2}}
\begin{tabular}{cccccccccc}
\hline
Frequency & ID & flux & distance& position angle & major\\
(GHz) & & [Jy] & [mas] & [${\circ}$]  & [mas] \\
\hline
\textbf{2014-09-01}&&&&&\\
22 & A  &  &  &  & \\ 
	 & B  &  0.26 &1.7 & -1.9 & 0.24 \\ 
	 & C  &  & &  &  \\\\  
	 & b1 & 0.016 & 9.0 & -1.0 & 0.91\\ 
	 & b2 & 0.043 & 9.6 & 6.7  & 1.1\\\\ 
\hline

\textbf{2014-09-14}&&&&&\\
22 & A  &  &  &  &  \\ 
	 & B  & 0.23 & 1.7 & -1.4& 0.30 \\ 
	 & C  &  &  & &  \\\\  
	 & b1 & 0.0064   & 9.2  & -2.2 & 0.30  \\ 
	 & b2 & 0.029   & 9.6  & 5.4  &  0.82 \\\\ 
\hline

\textbf{2014-11-04}&&&&&\\
22 & A  & 0.012 & 2.8 & -5.6 & 0.32 \\ 
	 & B  & 0.23   & 1.8 & -2.1& 0.32 \\ 
	 & C  &  &  &  &  \\\\  
	 & b1 &   0.027   &  9.0  &  -0.46   & 1.4 \\ 
	 & b2 &     0.028   &  9.8 &    5.7   & 0.91  \\\\ 
\hline

\textbf{2015-05-03}&&&&&\\
22 & A  &0.0090 & 3.1 & -2.2  & 0.38 \\ 
	 & B  & 0.19   &  1.9  &  -2.8   &  0.27 \\ 
	 & C  &      &  &  &  \\\\  
	 & b1 & 0.023   &   9.1 &-0.80  &  1.4 \\ 
	 & b2 &     0.032   &  9.8  &   6.7    & 1.1\\\\               
\textbf{2015-05-04}&&&&&\\
43 &	D &	0.091 & 2.2  & -1.2  & 0.41 \\
	 &	E &	0.068 & 0.73 & -1.4  & 0.24  \\
\hline

\textbf{2015-05-16}&&&&&\\
22 & A  & 0.0055 & 3.3 & -3.5 & 0.26 \\ 
	 & B  & 0.21   &  1.9  &  -2.8  & 0.36 \\ 
	 & C  & & & & \\\\  
	 & b1 & 0.024   &  9.1  &  -1.2   &  1.0 \\ 
	 & b2 & 0.031   &  9.8   &  6.1   &  0.97 \\\\ 
\textbf{2015-05-17}&&&&&\\
43 &	D &	0.10 & 2.2  & -1.3  & 0.46 \\
	 &	E &	0.090   &	 0.69 &  1.3  & 0.21  \\
\hline
\end{tabular}
\end{table}

\begin{table}
\centering
\contcaption{~}
\begin{tabular}{cccccccccc}
\hline
Frequency & ID & flux & distance& position angle & major\\
(GHz) & & [Jy] & [mas] & [${\circ}$]  & [mas] \\
\hline
\textbf{2016-02-25}&&&&&\\
22 & A  & 0.0036 & 3.8 & 1.7 & 0.38 \\ 
	 & B  & 0.12      & 2.5 &-1.2& 0.42\\ 
	 & C  & 0.44  & 0.52 &  13 & 0.055 \\\\  
	 & b1 &0.023    &  9.4  &  -0.71   &  1.3 \\ 
	 & b2 & 0.030   & 10   &  6.4   &  1.2 \\\\ 
\textbf{2016-02-26}&&&&&\\
43 &	D &	0.083 & 2.5  & 0.63  & 0.57 \\
	 &	E &	0.15   &	 0.86 & 8.7  & 0.16  \\
\hline

\textbf{2016-03-09}&&&&&\\
22 & A  & 0.0020 &   4.1&  6.6 & 0.39 \\ 
	 & B  & 0.12  &2.5  & -1.1  & 0.47 \\ 
	 & C  & 0.44  & 0.52 & 13  & 0.055\\\\  
	 & b1 & 0.019  &  9.5  & -0.55  &  1.2  \\ 
	 & b2 & 0.030  &  10 &  7.0 &1.2  \\\\ 
\textbf{2016-03-10}&&&&&\\
43 &	D &	0.078&  2.5 &  0.31 &0.50 \\
	 &	E &	 0.14 & 0.89 &  6.8  & 0.085  \\
\hline

\textbf{2016-03-20}&&&&&\\
43 &	D & 0.087 &  2.4 & 0.72 & 0.57 \\
	 &	E &0.16 & 0.87 & 9.1 & 0.041  \\
\textbf{2016-03-21}&&&&&\\
22 & A  & & & &  \\ 
	 & B  & 0.17  & 2.3 & -1.4& 0.50 \\ 
	 & C  &  0.40 & 0.74 & 8.3 & 0.048 \\\\  
	 & b1 &  0.027 &  9.5 &  1.0 &  0.92 \\ 
	 & b2 &  0.028 &  10 &  7.9& 0.67 \\\\ 
\hline

\textbf{2016-04-08}&&&&&\\
22 & A  & 0.0028  & 3.9 & 1.8 & 0.056 \\ 
	 & B  &  0.12 & 2.6& -1.1 & 0.44 \\ 
	 & C  &  0.41 & 0.57 & 13  & 0.007\\\\  
	 & b1 & 0.021  & 9.5 & -0.55 & 1.1\\ 
	 & b2 & 0.030  &  10  &  6.5  & 1.2\\\\ 
\textbf{2016-04-09}&&&&&\\
43 &	D &	0.085 &  2.5 &  -0.4 & 0.57 \\
	 &	E &	0.15&  0.90 & 8.1 & 0.090 \\
\hline

\textbf{2016-04-21}&&&&&\\
22 & A  &0.0058 & 3.9 & 5.2 & 0.43 \\ 
	 & B  & 0.10 & 2.6 & -0.67 & 0.38 \\ 
	 & C  &   0.38  & 0.60 & 13 & 0.023 \\\\  
	 & b1 & 0.033 &   9.6 &  2.6 &  1.7 \\ 
	 & b2 & 0.021 & 10.5 & 8.0 & 0.84  \\\\ 
\textbf{2016-04-22}&&&&&\\
43 &	D &	0.081 & 2.5  & 0.5  & 0.58 \\
	 &	E &	0.14   &	 0.90 & 9.9  & 0.16  \\
\hline
\end{tabular}
\end{table}

\begin{table}
\centering
\contcaption{~}
\begin{tabular}{cccccccccc}
\hline
Frequency & ID & flux & distance& position angle & major\\
(GHz) & & [Jy] & [mas] & [${\circ}$]  & [mas] \\
\hline 
\textbf{2016-05-03}&&&&&\\
22 & A  &&&& \\
	& B & 0.075 & 2.6 &  -1.1 & 0.34\\
	& C &   0.30 & 0.72 & 8.0  &  0.071\\
	& b1& 0.018 &  9.6 & 0.58 & 0.95\\
	& b2& 0.023 & 10 & 7.0  & 1.0\\\\
\textbf{2016-05-05}&&&&&\\
43 &	D & 0.056 & 2.6 & 2.6 & 0.63 \\
	 &	E & 	 0.14 & 0.84 & 14 & 0.16\\
\hline
\textbf{2016-05-23}&&&&&\\
22 & A & 0.0042 & 3.9 & 1.4  & 0.059 \\
	& B &  0.12 & 2.6 & -1.6 & 0.47\\
	& C&   0.33 & 0.59 & 14 & 0.073\\
	& b1& 0.024  &   9.6   &  0.078   &  1.1  \\
	& b2&  0.025   & 10  &    6.9&     1.0\\\\ 
\textbf{2016-05-24}&&&&&\\
43 &	D & 0.076 &    2.5 &   -0.053  &   0.60 \\
	 &	E &	  0.14  & 0.93  &   9.0 &   0.18 \\
\hline

\textbf{2016-06-02}&&&&&\\
43 &	D &	0.070 &   2.6 &   3.1 &   0.53 \\
	 &	E &  0.12  &  1.0  & 11 & 0.059 \\
\hline

\textbf{2016-06-13}&&&&&\\
22 & A &&&&\\
	& B & 0.056     & 2.7  & -1.7   &  0.15\\
	& C & 0.22 &    0.69 &    13 &     0.026\\
	& b1 & 0.011   &  9.8  &  0.79   & 0.42\\
	& b2 & 0.011  &  10   &    7.8  &     0.33 \\\\ 
\textbf{2016-06-15}&&&&&\\
43 &	D &	0.052   & 2.6   &  1.4 &  0.42  \\
	 &	E &  0.13 &   0.89 &    16 &    0.005   \\
\hline	
\end{tabular}
\end{table}

\bsp	
\label{lastpage}
\end{document}